\documentclass{iopart}

\usepackage{graphicx}
\begin{document}

\title{High-resolution saturation spectroscopy of singly-ionized iron with a pulsed uv laser}

\author{M Ascoli$^1$, E E Eyler$^1$, D Kawall$^{2,\footnotemark}$ and David DeMille$^2$}
\footnotetext{Present address: Department of Physics, University of Massachusetts, Amherst, MA, 01003, USA.}
\address{$^1$Physics Department,
University of Connecticut
Storrs, CT 06269, USA}
\address{$^2$Department of Physics,
Yale University
New Haven, CT, 06520, USA}
\ead{eyler@phys.uconn.edu}
\begin{abstract}
We describe the design and realization of a scheme for uv laser spectroscopy of singly-ionized iron (Fe II) with very high resolution. A buffer-gas cooled laser ablation source is used to provide a plasma close to room temperature with a high density of Fe II. We combine this with a scheme for pulsed-laser saturation spectroscopy to yield sub-Doppler resolution.  In a demonstration experiment, we have examined an Fe II transition near 260 nm, attaining a linewidth of about 250 MHz.  The method is well-suited to measuring transition frequencies and hyperfine structure.  It could also be used to measure small isotope shifts in isotope-enriched samples.
\end{abstract}

\maketitle

\section{Introduction}
\normalsize

There has recently been increased interest in precise spectroscopic data on the ultraviolet spectrum of singly-ionized metallic ions because of their importance as probes of the interstellar medium.  In particular, precise wavelengths and accurate isotope shift data are needed because of their impact on the analysis of quasar absorption systems.  Some recent studies indicate evidence for a space-time variation in the fine-structure constant $\alpha$ \cite {Murphy01,Webb01,Murphy03b,Flambaum07}, while others indicate a null effect \cite{Aracil04,Chand06,Levshakov06}, and there has been considerable effort to explain these discrepancies.  One potential systematic problem is that the isotope shifts for many of the key transitions are poorly known, and the cosmological evolution of isotope abundances could mimic an apparent change in $\alpha$ \cite{Murphy03,Kozlov04}.  In addition, some transition wavelengths are not known with sufficient accuracy to fully exploit the available astrophysical data.  A list of the spectral lines for which new laboratory data is needed most urgently has been published online \cite{Berengut06}, and includes numerous transitions in singly-ionized Mg, Ti, Mn, Fe, Ni, and other metals.  New theoretical calculations of specific isotope shifts in these species have been undertaken \cite{Berengut04,Berengut05}, but experimental confirmation is essential.

To help meet the need for new measurements, we demonstrate a new method for obtaining uv spectra of singly-ionized metallic atoms with sub-Doppler resolution, by the use of saturation spectroscopy with the harmonics of a pulse-amplified cw laser.  Previously saturation spectroscopy has been employed only rarely with pulsed lasers \cite{Hansch71,Hansch74}, and even then only on neutral atoms using visible lasers.  To provide a versatile source of cold ions suitable for pulsed laser excitation, we employ a laser ablation source incorporating buffer-gas cooling, and we show that the resulting plasma is very close to room temperature while remaining optically dense.

Compared with other spectroscopic techniques our approach has both advantages and disadvantages.  High-resolution Fourier transform spectroscopy can rapidly provide spectral data over large wavelength regions, with excellent signal to noise ratios and absolute accuracies as good as 0.001 cm$^{-1}$ \cite{Nave91,Aldenius06}.  However, the resolution is inherently limited by Doppler broadening, typically to about 0.2 cm$^{-1}$ for UV wavelengths near 250 nm.  This is usually insufficient to measure isotope shifts in metallic ions, which are frequently less than 0.1 cm$^{-1}$.  By contrast, our saturation spectroscopy scheme offers very high resolution, easily sufficient to resolve hyperfine structure or isotope shifts.  However, each spectral line must be scanned separately, and the signal to noise ratio is limited by shot-to-shot fluctuations in the ablation source and the pulsed spectroscopy laser.  For the case of Fe II, we find that the signal sizes are not quite adequate to unambiguously observe lines from the rare isotopes unless an isotopically enriched sample is used.

As with Fourier spectrometers, Doppler broadening limits the capabilities of laser spectroscopy using velocity modulation in discharges, which is otherwise appealing because of the the inherently large signal to noise ratio \cite{Oka03}.  It would also be difficult to apply this method to UV spectroscopy, because cw lasers are required, and cw harmonic generation is typically inefficient and lacks broad tunability.

There are alternative methods for obtaining very high resolution, but none with broad applicability.  Saturation spectroscopy of ions with cw lasers was demonstrated in ion beams many years ago \cite{Bayer79}, but the low ion beam flux severely limits the signal sizes, so the weak uv harmonics of a cw laser would be unusable, as would pulsed UV lasers with their low duty cycles.   In principle excitation of laser-cooled ions in an ion trap would be a nearly ideal scheme.  Here the long interaction time and low temperature permits the use of low-power uv harmonics of a cw laser.  This approach can yield superb results for favorable configurations \cite{Bergquist00}, and indeed it is the basis of the $^{199}$Hg$^+$ frequency standard, but neither the ion trap nor the lasers can easily be switched to other atomic species, and this is not presently a practical method for general spectroscopy of arbitrary ions.

For the initial results reported here we focus primarily on a strong absorption line in singly-ionized iron (Fe II), the $a\,^6D_{9/2} \rightarrow z\,^6D^o_{9/2}$ transition at 260.01 nm. The lower level of this strong transition is the ground state, and the $A$ coefficient for the upper level is $2.2 \times 10^8$ s$^{-1}$ \cite{NIST_tables}, corresponding to a natural linewidth of 35 MHz.  This line appears on the list of urgently needed data in Ref. \cite{Berengut06} because its isotope shifts are unknown, but its wavelength has been measured twice in recent years by UV Fourier transform spectroscopy \cite{Nave91,Aldenius06}.

\section{Experiment}

\subsection{Cold Plasma production}

To produce a relatively cold ion sample suitable for pulsed-laser spectroscopy, we have developed a laser ablation source that includes a buffer gas to cool the plasma plume resulting from the ablation pulse.  There is a large literature on the design and performance of laser ablation ion sources \cite{Hughes80,Lash96,Torrisi01,Nassisi03,Qi03,Amoruso02}, although in most cases the ions are much too energetic (tens to thousands of eV) for use in precision spectroscopy.  We were encouraged in to push ahead with the use of ablation together with a buffer gas by the results of a recent experiment on MgB$_2$ \cite{Amoruso02}, in which a plasma was cooled nearly to room temperature within a distance of a few mm by about 1 Torr of argon buffer gas.

\begin{figure}
\centering \vskip 0 mm
\includegraphics[width=0.8\linewidth]{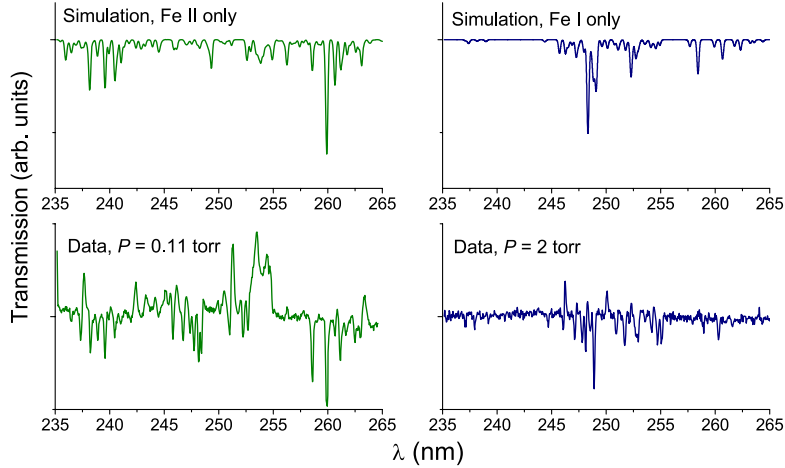}
\caption{\protect\label{fig:LowResSpectra} Simulated and measured low-resolution spectra.  Below 240 nm and near 260 nm, most of the lines are due to Fe II, and are greatly enhanced at 0.11 torr compared with 2 torr.  From 245-250 nm, most of the lines are due to Fe I.}
\end{figure}

For our laser ablation source, we use a dedicated 532 nm Nd:YAG laser with a typical pulse duration of 6-8 ns and an energy of 20-30 mJ.  It is focused onto an iron rod 5 cm in diameter, which is slowly moved in a spiral by a motor assembly to avoid laser drilling of the target.  The buffer gas is argon, and its pressure is stabilized by a computer-based ``proportional-integral-differential'' (PID) controller.

We optimize the ablation laser and the argon pressure by monitoring a low-resolution spectrum in the region 235-265 nm with both Fe I and Fe II transitions, obtained using a continuum light source and a monochromator.  As shown in \Fref{fig:LowResSpectra}, numerous transitions are seen both in absorption and emission.  To identify them we generate synthetic spectra using the wavelengths and intensities from the NIST atomic spectral data base \cite{NIST_tables}.  The intensity match is poor, in part because our iron spectra are observed mainly in absorption, while the NIST intensities reflect emission spectra.  Nevertheless it is possible to identify numerous Fe I and Fe II transitions.

\begin{figure}
\centering \vskip 0 mm
\includegraphics[width=0.6\linewidth]{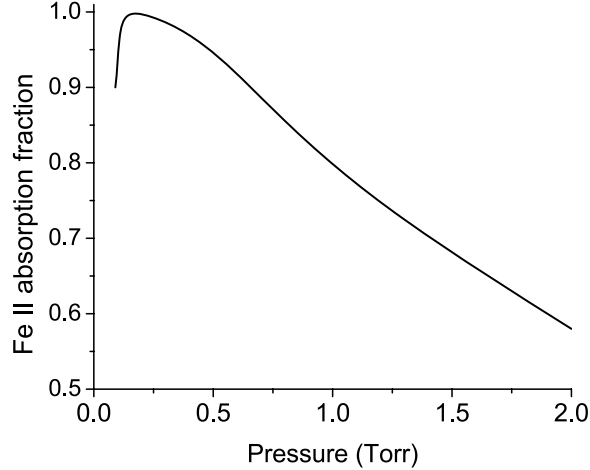}
\caption{\protect\label{fig:IonNeutralRatio} Absorption fraction for Fe II as a function of pressure, estimated by comparing Fe II to Fe I absorption line strengths in selected spectral regions (see text).}
\end{figure}

We also use these low-resolution spectra to optimize the buffer gas pressure for ground-state Fe II production in the plasma.  Although we have not attempted to make absolute density measurements, we use the relative line intensities of Fe I and Fe II to maximize the Fe II: Fe I ratio.  To quantify this, line strengths in the synthetic spectrum are compared to measured absorption spectra in three regions: 237-242 nm, containing mainly Fe II lines, 257-262 nm, containing the 260 nm Fe II line that we use for saturation spectroscopy, and 247-250 nm, containing mainly Fe I lines. An ``absorption fraction'' $f$ is determined by scaling the Fe I and Fe II synthetic spectra to produce a combined signal $S$ that best matches the data:
\begin{equation}
S_{\rm{total}} = A(fS_{\rm{Fe\ II}}  + (1 - f)S_{\rm{Fe\ I}} ).
\end{equation}
In \Fref{fig:IonNeutralRatio} we plot the pressure dependence of this fraction, which gives a rough indication of the fraction of the absorption line strength attributable to Fe II.  It exhibits a maximum for pressures near 0.2 Torr.

\subsection{Doppler and Sub-Doppler Laser Spectroscopy at 260 nm}

Our laser system starts with a cw ring Ti:sapphire laser operating near 780 nm, which is amplified into intense pulses by a pulsed dye amplifier, pumped by 9 nsec pulses from a 532 nm Nd:YAG laser.  The repetition rate is 10 Hz. The pulsed light is frequency doubled in a nonlinear crystal, then the fundamental is mixed with the second harmonic to produce the third harmonic at 260 nm.  These frequency-tripled pulses typically have bandwidths of about 100-200 MHz that can be close to the Fourier transform limit, as we have previously demonstrated \cite{Bergeson00,Yiannopoulou06}.  In the present case a transverse dye amplifier is used rather than a longitudinal capillary-type amplifier, so the bandwidth is probably close to the upper end of the 100-200 MHz range.  Although the laser system can easily produce pulse energies in excess of 200 $\mu$J at 260 nm, the output is attenuated to much lower energies for the measurements reported here.  To allow variation of the time delay between ablation and spectroscopy, separate Nd:YAG lasers were used for the two processes.  This delay is typically set to 400${\mu}$s.

\begin{figure}
\centering \vskip 0 mm
\includegraphics[width=0.6\linewidth]{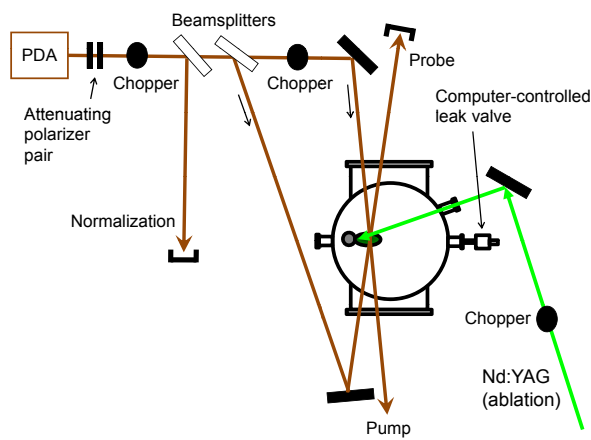}
\caption{\protect\label{fig:ExptScheme} Scheme for pulsed-laser saturation spectroscopy in the buffer gas-cooled iron plasma.  ``PDA'' denotes the pulsed dye amplifier}
\end{figure}

To calibrate the spectra, absolute frequency measurements are obtained from a commercial wavemeter (Burleigh WA-1500), accurate to 1 part in 10$^6$.  For more accurate calibrations, we make use of a heated I$_2$ absorption cell and an accurate synthetic spectrum created by the group of Tiemann \cite{Tiemann04}.  A temperature-stabilized confocal Fabry-Perot interferometer (Burleigh CF-25) provides marker fringes for scan linearization.

The layout of our saturation spectroscopy scheme is shown in \Fref{fig:ExptScheme}.  Counterpropagating pump and probe beams pass through the high-density region of the ablation plasma.  On strong absorption lines such as the 260 nm line, the ablation source can be adjusted to provide 80-90\% absorption of resonant light, making it easier to contend with shot-to-shot fluctuations of the pulsed laser.  Higher plasma densities are easily attainable, but are not helpful because the lasers are absorbed so strongly that signal analysis becomes difficult because of strong nonlinearities.

\begin{figure}
\centering \vskip 0 mm
\includegraphics[width=0.6\linewidth]{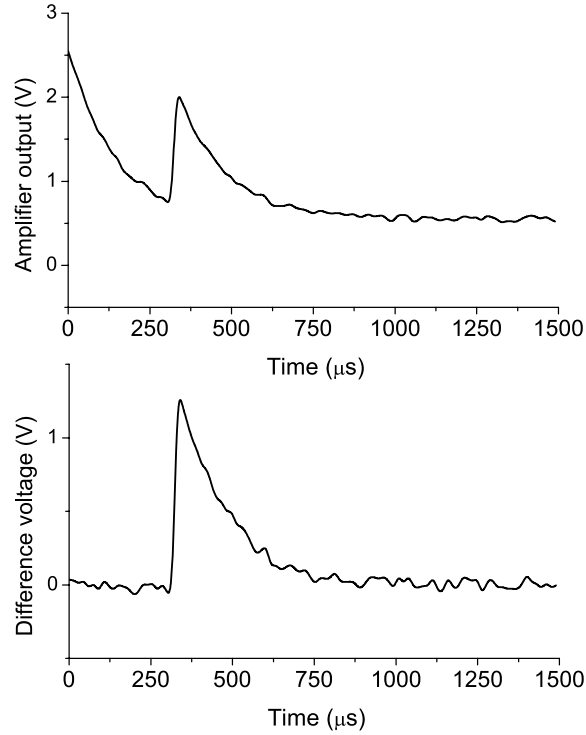}
\caption{\protect\label{fig:PreampVoltage} Top: output voltage from the charge-sensitive preamplifier after a single laser pulse.  Bottom: same, after subtracting background spectrum obtained with uv laser off.}
\end{figure}

This scheme differs from a classic three-beam cw saturation spectrometer \cite{Levenson88} by several adaptations to accommodate the pulsed laser and plasma source, which are subject to shot-to-shot fluctuations in addition to stray signals from optical and RF pickup.  To acquire a spectrum we scan the 260 nm laser in small steps of about 62 MHz.  At each step we sequentially measure four signals by using shutters to switch the two laser beams: (1) laser transmission, (2) pickup with all laser beams blocked, (3) Doppler-broadened absorption with the probe beam alone, and (4) the saturation signal, with pump and probe beams.  In addition we measure the laser power for each shot for normalization.  Each of these signals is acquired for several successive laser shots, after which the laser is stepped in frequency and the process is repeated.  After normalization, the residual amplitude fluctuations are on the order of 0.1\% per shot.  The signals are acquired using photodiodes with charge-sensitive preamplifiers.  In \Fref{fig:PreampVoltage} we show a typical output pulse, both before and after subtracting the background signal from the ablation laser.

By subtracting the single-beam Doppler signal from the counterpropagating-beam signal, we can obtain a pure saturation spectrum.  To observe saturation spectra without excessive broadening we greatly attenuate the lasers, to a probe energy of ~3 nJ/pulse (an irradiance of 1.5 W/cm$^2$) and a pump energy of ~53 nJ/pulse (28 W/cm$^2$, to be compared with an estimated cw saturation irradiance of 0.26 W/cm$^2$).

\section{Results}

When the absorption of a single uv laser beam by the plasma is monitored while scanning over an absorption feature, we see a Gaussian line profile with a full width at half maximum (FWHM) of 2.3 GHz using a buffer gas pressure of 0.2 torr.  If this is attributable entirely to Doppler broadening, as seems likely, the temperature of the ions is about 440 K.  This indicates that the collisional cooling is extremely effective, achieving a nearly room-temperature sample without causing excessive recombination of the ions.

\begin{figure}
\centering \vskip 0 mm
\includegraphics[width=0.6\linewidth]{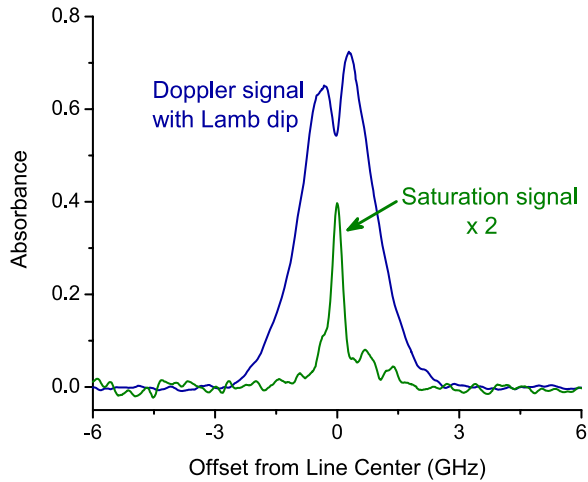}
\caption{\protect\label{fig:AvgSignal} Absorbance (ln$(I_o/I$)) of the $a\,^6D_{9/2} \rightarrow z\,^6D^o_{9/2}$ transition in Fe II at 38458.988 cm$^{-1}$.  Upper curve: Doppler-broadened pedestal and Lamb dip from pulsed-laser saturation spectroscopy.  Lower curve: Saturation signal after subtraction of Doppler-broadened spectrum, inverted for clarity.}
\end{figure}

We find that the highest-quality saturation spectra are obtained by acquiring several independent scans and averaging them, after aligning the frequency axis using the iodine spectrum.  In \Fref{fig:AvgSignal} we show a five-scan average after smoothing appropriate to the signal width.  The Doppler-broadened signal exhibits an obvious saturation dip.  The pure saturation signal is obtained by subtracting the signal due to the probe beam alone.  The Doppler-free full width at half maximum (FWHM) of the saturation feature is typically 250 MHz, which we believe to be the best resolution ever attained for the uv spectrum of an ion.  For example, it is 24 times narrower than the FWHM of the same transition in the high-resolution Fourier-transform spectra of Refs. \cite{Nave91,Aldenius06}.  The linewidth slightly exceeds the estimated laser bandwidth.  This is almost undoubtedly due to power broadening, because the pump laser irradiance used for the measurements is considerably in excess of the calculated saturation irradiance.

The small features to the right of the main saturation signal peak in \Fref{fig:AvgSignal} are probably due at least in part to the $^{54}$Fe isotope (5.8 \% abundance \cite{NNDC}), but the signal to noise ratio is not quite sufficient to be certain of their existence, so the isotope shift cannot be determined from this demonstration experiment.  If both peaks are real, one possibility is that they correspond to the $^{54}$Fe isotopic line and to a collision-induced crossover resonance \cite{Johns75}.  It is unlikely that $^{57}$Fe could have been observed in this experiment, because its abundance is only 2.1\% and its nuclear spin of 1/2 causes hyperfine splitting.  In principle the isotopic lines could easily be determined unambiguously by substituting an isotope-enriched iron sample, but limitations on time and resources prevented us from attempt this for the present experiment.

Even though this experiment did not include full provisions for precise frequency metrology, we are able to make a reasonably accurate determination of the center frequency by using the Doppler-broadened iodine absorption spectrum as a reference.  The principal uncertainties are statistical noise in the saturation and iodine spectra as well as random scatter due to nonlinearities within the individual laser scans, probably due to shifts in the Ti:sapphire laser as it is discontinuously stepped.  We have observed no statistically significant dependence on the laser power or the ablation conditions.  We have determined the line center for each of eight separate scans by fitting to Lorentzian lineshapes.  There is no hyperfine structure for $^{56}$Fe because of its spin-zero nucleus, and the line center is not affected appreciably by inclusion or exclusion of simultaneous fits to the two small features to the right of the main peak.  The average line center is 38458.988$\pm 0.004$ cm$^{-1}$, where the uncertainty is taken to be the standard deviation of the eight results (we do not divide by $\sqrt{8}$ because it is not clear how much of the scatter is systematic in nature, and because we have not controlled carefully for ac Stark shifts or collisional shifts).  This result is slightly less accurate but in excellent agreement with the most recent high-resolution Fourier spectrometer measurement \cite{Aldenius06}, 38458.991$\pm 0.002$ cm$^{-1}$.

\section{Conclusions}

We have demonstrated a scheme for sub-Doppler saturation spectroscopy of atomic ions in the mid-UV region, using a frequency-tripled pulse-amplified cw laser.  We have developed a simple and versatile ablation source with buffer gas cooling that can produce optically dense samples of Fe II close to room temperature.  The spectral resolution is at least 250 MHz, and might be further reduced by reducing the laser power.

Although the scheme is very well suited to measuring transition frequencies and closely-spaced hyperfine structure, its application to isotope shift measurements is limited by the signal to noise ratio of roughly 25:1.  This can readily be circumvented, however, by using isotope-enriched samples.

Because the pulse-amplified laser is broadly tunable, and the ion source will work with nearly any metallic species as well as many non-metals, we believe that the performance of this saturation spectrometer will be very similar for other atomic ions.  Extension to higher charge states is also a possibility, but has not yet been explored.

%Acknowledgements
\ack
We wish to acknowledge support from a National Science Foundation grant to the University of Connecticut and from Yale University.

\end{document}